\documentclass[%
 aip,
 amsmath,amssymb,
 reprint,%
]{revtex4-1}

\usepackage{graphicx}
\usepackage{dcolumn}
\usepackage{bm}

\usepackage[utf8]{inputenc}
\usepackage[T1]{fontenc}
\usepackage{mathptmx}
\usepackage{etoolbox}

\graphicspath{ {./FIGS} }

\makeatletter
\def\@email#1#2{%
 \endgroup
 \patchcmd{\titleblock@produce}
  {\frontmatter@RRAPformat}
  {\frontmatter@RRAPformat{\produce@RRAP{*#1\href{mailto:#2}{#2}}}\frontmatter@RRAPformat}
  {}{}
}%
\makeatother
\begin{document}

\preprint{AIP/123-QED}

\title{Control of the nonlinear frequency shift for the spin-transfer nanooscillator using a bias magnetic field}
\author{A.A Matveev}
 \affiliation{Kotel’nikov Institute of Radio-Engineering and Electronics of RAS, Moscow 125009, Russia}
 \affiliation{Moscow Institute of Physics and Technology (National Research University), Dolgoprudnyi, Moscow region, 141701 Russia}
\email{maa.box@yandex.ru}
\author{A.R. Safin}%
\affiliation{Kotel’nikov Institute of Radio-Engineering and Electronics of RAS, Moscow 125009, Russia}
\affiliation{National Research University Moscow Power Engineering Institute, Moscow, 111250 Russia}

\author{S.A. Nikitov}
 \affiliation{Kotel’nikov Institute of Radio-Engineering and Electronics of RAS, Moscow 125009, Russia}
 \affiliation{Moscow Institute of Physics and Technology (National Research University), Dolgoprudnyi, Moscow region, 141701 Russia}

\date{\today}

\begin{abstract}
We investigated the possibilities of controlling the nonlinear frequency shift of the magnetization oscillations in a spin-transfer nanoscillator by varying the magnitude and direction of the bias magnetic field. We considered both isotropic ferromagnetic materials and crystals with uniaxial and cubic crystallographic anisotropies. We have shown that achieving a zero nonlinear frequency shift is possible with a certain orientation of the bias magnetic field vector. The results of the theoretical analysis based on the method of Hamiltonian formalism are in good agreement with the micromagnetic simulations. Our research reveals the way to control the frequency tuning of a spin transfer nanoscillator, which is crucial for spintronic signal generation devices.
\end{abstract}

\maketitle

The possibility of manipulating magnetization using spin-polarized current enables the development of various spintronic devices \cite{Shao}. Based on the spin-torque effect, it is proposed to design magnetic memory, nanogenerators in gigahertz and terahertz frequency ranges, neuromorphic circuits, magnetic field sensors, amplifiers and other devices \cite{Shao, Zvez2020, Loc2014, Zhong2020, SUN2022, Zhu}. The main element in such devices is typically a spin-transfer nanoscillator (STNO) \cite{new_MTJ, MTJ_app, Pathak2020, Shao}. Such a structure is a stack of free and fixed layers separated by a spacer. The dynamics of magnetization in the free layer determines the properties of the STNO in spintronic devices \cite{MTJ_app}. The emergence of self-oscillations in a ferromagnetic free layer was theoretically described using the classical Hamiltonian formalism for spin waves \cite{Sl_Tib_star, Sl_Tib_ras}. It is known \cite{Magnon, Sl_Tib_ras, Rippard, Costa_self} that there is a certain critical density of the spin polarized current passed through the magnetic film at which the ground state of magnetization turns out to be unstable and supercritical bifurcation occurs. This happens because spin-transfer compensates the intrinsic damping of the magnetic material \cite{Costa_self}. However, with an increase in the density of the spin polarized current passed through the ferromagnet, the amplitude of the precession of magnetization increases \cite{Sl_Tib_ras}. The dependence of the frequency of self-oscillations on the density of such a current turns out to be more complicated. It can be affected by the bias magnetic field \cite{Sl_Tib_star}. With growth of the density of the spin-polarized current, the frequency of the STNO can both increase and decrease \cite{Sl_Tib_ras}. In this paper, we investigate the \textit{impact of the bias magnetic field on the frequency shift of the magnetization precession in the STNO's free layer}. Moreover, when applying the classical Hamiltonian formalism to the description of the dynamics of the structure under study, we take into account the existence of the magnetic crystallographic anisotropy in magnetic thin film. Previously\cite{Sl_Tib_ras, Sl_Tib_star}, using such a method, only isotropic magnetics and materials with uniaxial anisotropy were considered. 
\begin{figure}
\includegraphics[scale=0.75]{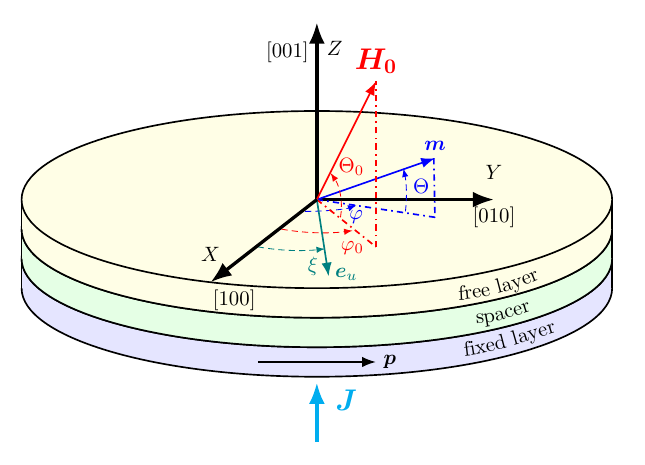}
\caption{\label{fig:FIG_1} Scheme of the investigated system. The directions of the axes of the chosen Cartesian coordinate system coincide with the vectors of cubic crystallographic anisotropy. The direction of the injected spin-polarized current is indicated by a cyan arrow. The polarization vector of a fixed layer is denoted as $\mathbf{p}$. The angles $\Theta_0,\ \varphi_0$ set the direction of the bias magnetic field $\mathbf{H}_0$. The vector of uniaxial anisotropy $\mathbf{e}_u$ lies in the plane of free layer and is oriented in it by an angle $\xi$. The angles $\Theta,\ \varphi$ set the direction of the magnetization vector $\mathbf{m}$.}
\end{figure}

The investigated system (see Fig.~\ref{fig:FIG_1}) is the STNO. It consists of a fixed magnetic layer that serves as a polarizer, a nonmagnetic separator (spacer) and a ferromagnetic free layer \cite{STNOs}. The direct current injected into the structure spin-polarizes when passing through a fixed layer \cite{MTJ_app}. The spin-polarized current  creates a spin-transfer torque in the free layer \cite{new_MTJ, STT}. This spin-transfer torque creates negative magnetic damping capable of compensating the internal positive damping of the ferromagnet. When spin-polarized current density is large enough, the self-oscillatory precession of magnetization begins. That is, supercritical bifurcation occurs and the magnetization oscillates near the magnetic energy minimum. With a further rise in the density of the spin-polarized current, the amplitude of the magnetization oscillations increases. However, with a sufficiently large current, the equilibrium state of magnetization around which the dynamics occurs can become an energy maximum. Previously\cite{Sl_Tib_ras, Sl_Tib_star}, this behavior of magnetization in a free layer was studied by the method of Hamiltonian formalism. Nevertheless, the study of the possibilities of controlling a nonlinear frequency shift using the bias magnetic field magnitude and orientation has not been carried out. With the help of the magnetoresistive effect, the dynamics of magnetization in the free layer is converted into a microwave voltage \cite{STNOs, Grimaldi}. Thus an electrical signal is generated. The frequency of this ac signal coincides with the frequency of the magnetization precession. That is why it is crucial to be able to adjust the frequency of magnetization oscillations in order to design spintronic signal generation devices.

The investigated structure (see Fig.~\ref{fig:FIG_1}) is subjected to the bias magnetic field $\mathbf{H}_0$. The dynamics of magnetization in this case can be described using the Landau-Lifshitz equation\cite{Atxitia_2017, Sl_Tib_ras}
\begin{equation}\label{eq:01}
    \dot{\mathbf{m}}=-\gamma\ \mathbf{m}\times\mathbf{H}_{eff}+\alpha\ \mathbf{m}\times\dot{\mathbf{m}}+\beta J\ \mathbf{m}\times\left[\mathbf{p}\times\mathbf{m}\right].
\end{equation}
Here, the first term with effective magnetic field $\mathbf{H}_{eff}$ describes all the self-interactions and the influence of the bias magnetic field (see below), the second term takes into account the Gilbert damping\cite{G_damping} with constant $\alpha$, the last term arises due to the flow of a spin-polarized current with a density $J$ through the free layer. In equation (\ref{eq:01}) $\gamma$ is the module of the gyromagnetic ratio of the electron, $\beta=\gamma\hbar P/\left(2 M_s e L\right)$ is the normalization constant for the current, $\hbar$ is the reduced Planck constant, P is the degree of polarization\cite{Polar}, $M_s$ is the saturation magnetization, $e$ is the elementary electric charge, $L$ is the free layer thickness. The effective magnetic field is defined as the variational derivative of the magnetic energy functional $E$ \cite{Sl_Tib_ras, Sl_Tib_star}
\begin{equation}\label{eq:02}
    \mathbf{H}_{eff}=-\frac{1}{\mu_0M_s}\frac{\delta E}{\delta \mathbf{m}},
\end{equation}
where $\mu_0$ is the vacuum magnetic permeability. In the chosen coordinate system (see Fig.~\ref{fig:FIG_1}) with macrospin approximation, we can consider the magnetic energy in the form
\begin{eqnarray}\label{eq:03}
  E&=&\int\biggr{(}-\mu_0M_s\mathbf{H}_0\cdot\mathbf{m}+\frac{\mu_0M_s^2}{2}\left(\mathbf{m}\cdot\mathbf{e}_z\right)^2\nonumber\\ 
  &-&K_u\left(\mathbf{m}\cdot\mathbf{e}_u\right)^2-\frac{K_c}{2}\left(m_x^4+m_y^4+m_z^4\right)\biggr{)}\ dV.
\end{eqnarray}
Here the first term represents the Zeeman energy associated with the application of a magnetic field $\mathbf{H}_0$, the second term is the shape anisotropy (demagnetization energy) of a thin film arising from the dipole-dipole interaction in a ferromagnet, the third term takes into account uniaxial magnetic anisotropy and the last term arises due to the cubic crystallographic anisotropy. The anisotropy constants $K_u,\ K_c$ are connected to the frequently used anisotropy fields $H_u,\ H_c$ through the relations $H_u=2K_u/\left(\mu_0M_s\right),\ H_c=2K_c/\left(\mu_0M_s\right)$. 

The core of the Hamiltonian formalism\cite{Patton, Sl_Tib_star, Lvov} is the transformation of the Landau-Lifshitz equation to the Hamiltonian equations. To achieve this, it is necessary to introduce a new variable describing the dynamics of magnetization. When the current density exceeds a certain critical value in the free layer, an auto-oscillatory precession of magnetization begins. Such magnetization dynamics are often characterized by the amplitude and frequency of precession \cite{STNOs, MTJ_app, Sl_Tib_ras}. Therefore, one of the simplest methods of introducing a new variable is a transformation that connect this new variable with the amplitude of precession. Hence, first we look for the direction around which the magnetization is precessing. We consider a case when the current density is not too high and the magnetization oscillates around the minimum energy position of the vector $\mathbf{m}$ \cite{Sl_Tib_star, Sl_Tib_ras}. The ort $\mathbf{e}_{\zeta}$ of such direction is found from the equation 
\begin{equation}\label{eq:04}
    -\frac{1}{\mu_0M_s}\frac{\delta E}{\delta \mathbf{m}}=H \mathbf{e}_{\zeta},
\end{equation}
where $H$ is an internal magnetic field. In the Cartesian coordinate system, the vector $\mathbf{e}_{\zeta}$ is represented as $\mathbf{e}_{\zeta}=\left(\cos\left(\Theta\right)\cos\left(\varphi\right), \cos\left(\Theta\right)\sin\left(\varphi\right), \sin\left(\Theta\right)\right)$. Then the orientation angles of the vector $\mathbf{e}_{\zeta}$ and the internal magnetic field $H$ can be found from the system of equations
\begin{eqnarray}\label{eq:05}
 &H&\cos\left(\Theta\right)\cos\left(\varphi\right)=H_0\cos\left(\Theta_0\right)\cos\left(\varphi_0\right)\nonumber\\
 &+&H_u \left(\mathbf{e}_{\zeta}\cdot\mathbf{e}_{u}\right)\cos\left(\xi\right)+Hc\left(\mathbf{e}_{\zeta}\cdot\mathbf{e}_{x}\right)^3, \nonumber\\[8pt]
&H&\cos\left(\Theta\right)\sin\left(\varphi\right)=H_0\cos\left(\Theta_0\right)\sin\left(\varphi_0\right)\\
 &+&H_u \left(\mathbf{e}_{\zeta}\cdot\mathbf{e}_{u}\right)\sin\left(\xi\right)+Hc\left(\mathbf{e}_{\zeta}\cdot\mathbf{e}_{y}\right)^3, \nonumber\\[8pt]
 &H&\sin\left(\Theta\right)=H_0\sin\left(\Theta_0\right)+H_c\left(\mathbf{e}_{\zeta}\cdot\mathbf{e}_{z}\right)^3.\nonumber
\end{eqnarray}We introduce two orts $\mathbf{e}_{\xi},\ \mathbf{e}_{\eta}$ that complement the $\mathbf{e}_{\zeta}$ to the right triplet

\begin{eqnarray}\label{eq:06}
\mathbf{e}_{\xi}&=&\sin\left(\Theta\right)\cos\left(\varphi\right)\mathbf{e}_x+\sin\left(\Theta\right)\sin\left(\varphi\right)\mathbf{e}_x-\cos\left(\Theta\right)\mathbf{e}_z,\nonumber \\
\mathbf{e}_{\eta}&=&-\sin\left(\varphi\right)\mathbf{e}_x+\cos\left(\varphi\right)\mathbf{e}_x.
\end{eqnarray}
Hamiltonian equations are written with respect to variables called canonical\cite{Sl_Tib_ras}. Such variables can be complex-conjugated and depend on coordinates and time. Hamiltonian equations are convenient for the analysis of nonlinear dynamics\cite{Patton} and, for example, are used in the study of plasma\cite{Plasma}, Bose-Einstein condensate\cite{ZAKHAROV_BE} and fluid dynamics\cite{LVOV_fluid}. When applied to the description of the dynamics of magnetization, canonical variables are most often associated with the deviation of the vector $\mathbf{m}$ from its equilibrium position \cite{Patton, Sl_Tib_ras}. For spin waves, the canonical transformation was first presented by Holstein and Primakoff\cite{Holt}. In our case it can be written as
\begin{equation}\label{eq:07}
   a=\frac{m_\xi-im_\eta}{\sqrt{2\left(1+m_\zeta\right)}}.
\end{equation}
Here $m_\xi,\ m_\eta,\ m_\zeta$ are the projections of the vector $\mathbf{m}$ on the corresponding coordinate axes. The dimensionless variable $a$ describes the amplitude of the precession of magnetization in the free layer. The dynamics equation for this variable has the quasi-Hamiltonian form
\begin{equation}\label{eq:08}
   \frac{\partial a}{\partial t}=-i  \frac{\partial \mathcal{H}_a}{\partial a^*}+F_a,
\end{equation}
where $\mathcal{H}_a=\gamma E/\left(\mu_0M_s\right)$ is the normalized energy that depends on the complex amplitude $a$, $F_a$ is the perturbation force. This force takes into account the influence of spin polarized current and Gilbert damping. The derivation of the expression for $F_a$ can be found in earlier works on the Hamiltonian formalism for spin-transfer oscillators\cite{Sl_Tib_ras}. Since the effect that we show in our work is related to the conservative part of the Hamiltonian equations for complex amplitudes, we do not represent the derivation of expressions for perturbation forces.

The energy $\mathcal{H}_a$ can be decomposed in a series by a dimensionless variable $a$ in the vicinity of the minimum magnetic energy. Then one can write
\begin{eqnarray}\label{eq:09}
\mathcal{H}_a&=&A|a|^2+\frac{1}{2}\left(Ba^2+B^*a^{*2}\right)+\left(V|a|^2a+V^*|a|^2a^*\right)\nonumber\\&+&U_1|a|^4+\left(U_2|a|^2a^2+U_2^*|a|^2a^{*2}\right)+\dots.
\end{eqnarray} Expressions for the coefficients of the series (\ref{eq:09}) are presented below
\begin{widetext}
\begin{eqnarray}\label{eq:A1}
&A&=\gamma\biggr{(}\left(H_u\left(\cos(\xi)^2-\frac{1}{2}\right)\cos(\varphi)^2-\frac{1}{2}H_u\cos(\xi)^2 +\frac{H_u}{4}\sin(2\xi)\sin(2\varphi)-3H_c-\frac{1}{2}Hu+\frac{1}{2}Ms\right)\cos(\Theta)^2\nonumber\\
&+&3H_c\left(\cos(\varphi)^4-\cos(\varphi)^2+1\right)\cos(\Theta)^4+H_0-\frac{1}{2}H_u\biggr{)}, \\[8pt]
&B& = \gamma\biggr{(}-\frac{1}{4}H_c\left(6\cos(\varphi)^2\cos(\Theta)^2(\cos(\varphi)\sin(\Theta)-i\sin(\varphi))^2+6\sin(\varphi)^2\cos(\Theta)^2(i\cos(\varphi)+\sin(\varphi)\sin(\Theta))^2+6\sin(\Theta)^2\cos(\Theta)^2\right)\nonumber\\
&+&\frac{1}{2}M_s\cos(\Theta)^2-H_u(\cos(\varphi)\sin(\Theta)\cos(\xi)+\sin(\varphi)\sin(\Theta)\sin(\xi)+i(\cos(\varphi)\sin(\xi)-\sin(\varphi)\cos(\xi)))^2
\biggr{)},\\[8pt]
&V&= -9\gamma\cos(\Theta)\biggr{(}\biggr{(}-H_c\cos(\varphi)^4\cos(\Theta)^2+\left(H_c\cos(\Theta)^2-\frac{2H_u}{9}\cos(\xi)^2+\frac{H_u}{9}\right)\cos(\varphi)^2-\frac{H_u}{18}\sin(2\xi)\sin(2\varphi)-H_c\cos(\Theta)^2 \nonumber\\
&+&\frac{H_u}{9}\cos(\xi)^2+\frac{H_c}{2}-\frac{H_u+Ms}{9}\biggr{)}\sin(\Theta)+iH_c\cos(\varphi)^3\cos(\Theta)^2\sin(\varphi)-\frac{iH_u}{9}\sin(2\xi)\cos(\varphi)^2\nonumber\\
&-&\frac{i}{2}\sin(\varphi)\left(H_c\cos(\Theta)^2-\frac{4H_u}{9}\cos(\xi)^2+\frac{2H_u}{9}\right)\cos(\varphi)+\frac{iH_u}{9}\cos(\xi)\sin(\xi)\biggr{)},\label{eq:12}\\[8pt]
&U_1&=-\frac{45\gamma}{2}\biggr{(}H_c(\cos(\varphi)^4-\cos(\varphi)^2+1)\cos(\Theta)^4+\frac{H_c}{5}-\frac{H_u+2Ms}{45}\nonumber\\
&+&\left(\frac{H_u}{15}(2\cos(\xi)^2-1)\cos(\varphi)^2+\frac{H_u}{30}\sin(2\xi)\sin(2\varphi)-\frac{H_u}{15}\cos(\xi)^2-\frac{H_u+Ms}{15}\right)\cos(\Theta)^2\biggr{)}\label{eq:13},\\[8pt]
&U_2&=-\frac{17\gamma}{64}\biggr{(}\frac{2H_u}{17}\cos(2\varphi-2\xi-2\Theta)+iH_c\cos(4\varphi-3\Theta)+\frac{2H_u}{17}\cos(2\Theta+2\varphi-2\xi)-iH_c\cos(3\Theta+4\varphi)+\frac{H_c}{4}\cos(4\varphi-4\Theta)\nonumber\\
&-&H_c\cos(4\varphi-2\Theta)-H_c\cos(4\varphi+2\Theta)+\frac{H_c}{4}\cos(4\varphi+4\Theta)+\frac{8iH_u}{17}\cos(2\xi-\Theta+2\varphi)-\frac{8iH_u}{17}\cos(\Theta-2\xi+2\varphi)+iH_c\cos(4\varphi-\Theta)\nonumber\\
&-&\frac{12H_u}{17}\cos(2\varphi-2\xi)-iH_c\cos(\Theta+4\varphi)+\frac{34H_c+4H_u+8M_s}{17}\cos(2\Theta)+\frac{7H_c}{2}\cos(4\Theta)-\frac{5H_c}{2}cos(4\varphi)-\frac{3Hc}{2}+\frac{4H_u+8M_s}{17}\biggr{)}.\label{eq:14}
\end{eqnarray}
\end{widetext}
The obtained expressions for the coefficients of Hamiltonian cover a wide range of magnetic materials. They are suitable for isotropic ferromagnets for the description of which it is necessary to set $H_u=0,\ H_c=0$ and for crystals in which there are uniaxial and cubic crystallographic anisotropies.
The Hamiltonian obtained (\ref{eq:09}) is quite cumbersome for analysis. Therefore, some transformations should be applied. Let's start with the Bogolyubov's diagonalization\cite{Bogoljubov}. This transformation simplify the quadratic part of the Hamiltonian\cite{Sl_Tib_ras, Lvov}. In the investigated case, it can be written as  
\begin{equation}\label{eq:15}
   a=ub-vb^*,
\end{equation}
where
\begin{eqnarray}\label{eq:16}
&u&=\sqrt{\frac{A+\omega_0}{\omega_0}},\nonumber\\
&v&=\sqrt{\frac{A-\omega_0}{\omega_0}},\\
&\omega_0&=\sqrt{A^2-\left|B\right|^2}.\nonumber
\end{eqnarray}
With such a substitution, the decomposition for energy has the form
\begin{eqnarray}\label{eq:17}
\mathcal{H}_b&=&\omega_0|b|^2+\left(W_1|b|^2b+W_1^*|b|^2b^*+W_1b^3+W_1^*b^{*3}\right)\nonumber\\
&+&T|b|^4+\dots.
\end{eqnarray}
Here coefficients are given by 
\begin{eqnarray}\label{eq:18}
W_1&=&-v^{*2}V^{*}v+2u(Vv-u^*V^*)v^*+u^*Vu^2,\\
W_2&=&V^{*}v^{*2}-Vu^2v^*,\label{eq:19}\\ 
T&=&(u^2U_1-3uU_2^*v^*)u^{*2}-3(u^2U_2-\frac{4}{3}uU_1v^*+v^{*2}U_2^*)vu^*\nonumber\\
&-&3v^*U_2uv^2+v^{*2}U_1v^2.\label{eq:20}
\end{eqnarray}
Note that diagonalization (\ref{eq:15}) is a canonical transformation that is it preserves the Hamiltonian form of the equations. Then the conservative dynamics equation for the complex amplitude $b$ can be written as
\begin{equation}\label{eq:21}
   \frac{\partial b}{\partial t}=-i  \frac{\partial \mathcal{H}_b}{\partial b^*}.
\end{equation}
It is possible to eliminate cubic terms in the (\ref{eq:17}). However, to do this, one have to use a quasi-canonical nonlinear substitution. This means that the Hamiltonian form will be preserved only if the high-order terms are discarded in the decomposition of the Hamiltonian. As it was shown previously\cite{Sl_Tib_ras, Lvov}, following nonlinear substitution can be used
\begin{equation}\label{eq:22}
   b=c+\frac{1}{\omega_0}\left(W_1c^2-2W_1^*|c|^2-W_2^*c^{*2}\right).
\end{equation}
With such a transformation, the Hamiltonian takes the simple form
\begin{equation}\label{eq:23}
   \mathcal{H}_c=\omega_0|c|^2+\frac{N}{2}|c|^4,
\end{equation}
where the non-isochronicity coefficient is given by
\begin{equation}\label{eq:24}
   N=2\left(T-3\frac{\left|W_1\right|^2+\left|W_2\right|^2}{\omega_0}\right).
\end{equation}
 This coefficient determines the nonlinear frequency shift of the self-oscillatory precession of magnetization in accordance with the expression\cite{Sl_Tib_star}
\begin{equation}\label{eq:25}
   \Omega=\omega_0+Np.
\end{equation}
Here $\Omega$ is the magnetization oscillations frequency, $p=|c|^2$ is the power of the precession. If the self-oscillatory precession of magnetization occurs near the minimum energy, then the power always increases with the increase in the density of the spin-polarized current. Then $N$ determines how the frequency depends on the density of the spin-polarized current passed through the free layer. Since the coefficient $N$  can be adjusted using a bias magnetic field, it is possible to control the nonlinear frequency shift of the STNO.

To verify the theoretically obtained result, micromagnetic simulations was carried out using $MUMAX^3$ software\cite{mumax3}. We have modeled both an isotropic material $\text{Ni}_{80}\text{Fe}_{20}$ and a material $\text{La}_{0.7}\text{Sr}_{0.3}\text{Mn}\text{O}_3$ where uniaxial and cubic crystallographic anisotropies are present. For the first isotropic\cite{Mat1} $\text{Ni}_{80}\text{Fe}_{20}$ case the following parameters were used $M_s=8\cdot10^5A/m$, $\alpha=0.01$, the  exchange stiffness $A_{ex}=13\cdot10^{-12} J/m$. The following parameters were selected for modeling a non-isotropic\cite{Mat2} $\text{La}_{0.7}\text{Sr}_{0.3}\text{Mn}\text{O}_3$ material $M_s=3\cdot10^5A/m$, $\alpha=1.9\cdot10^{-3}$, $A_{ex}=1.7\cdot10^{-12} J/m$, $H_u=8.4\cdot10^3 A/m$, $H_c=1.12\cdot10^3 A/m$, $\xi=\pi/4$. In both cases an orthogonal parallelepiped with periodic boundary conditions with dimensions $50\times50\times5 \text{nm}^3$ divided into $32\times32\times4$ simulation cells was modeled.
\begin{figure}[b]
\includegraphics[scale=0.5]{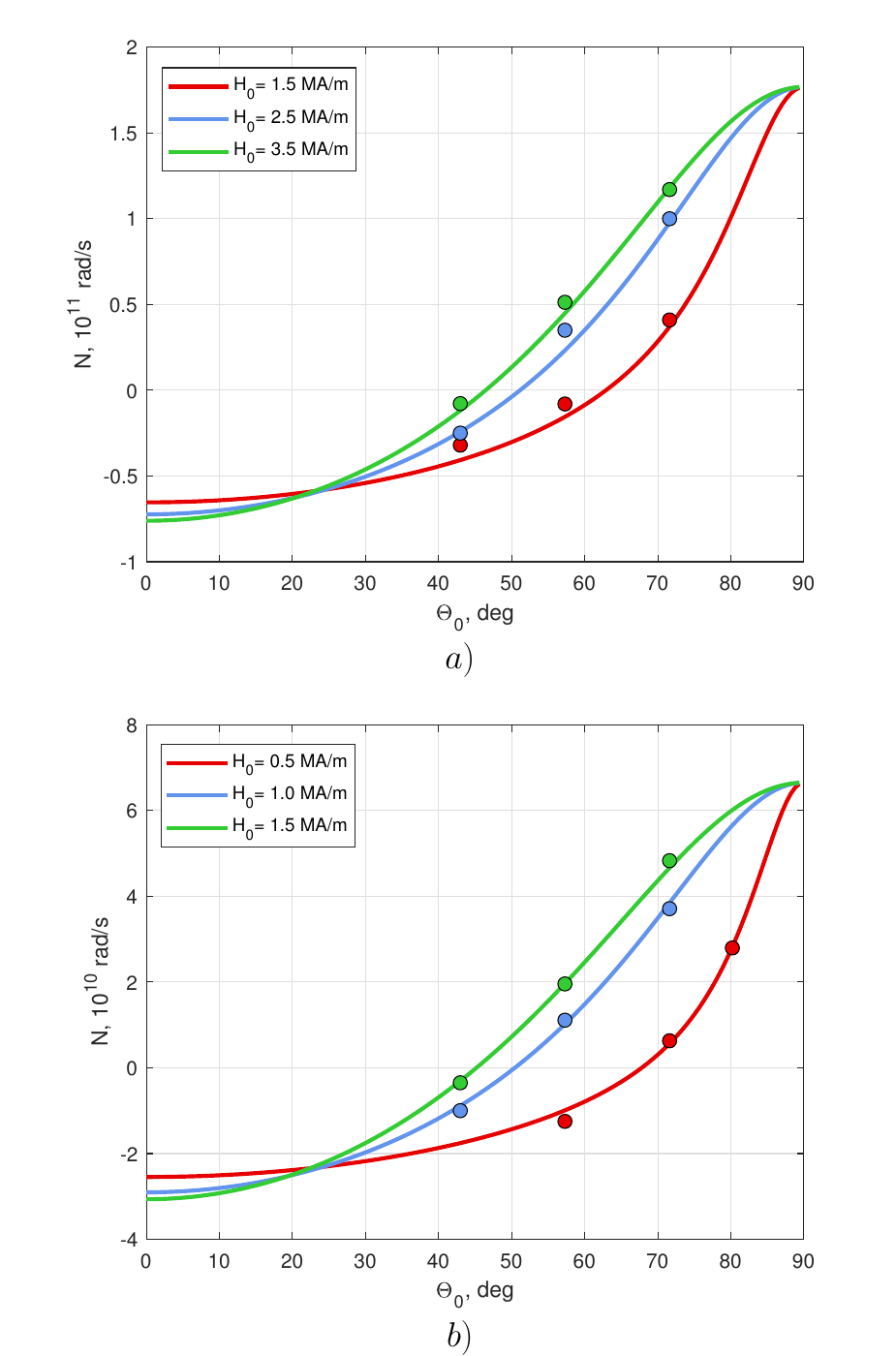}
\caption{\label{fig:FIG_2} The coefficient $N$ as a function of the polar angle $\Theta_0$ for the free layer made of a) $\text{Ni}_{80}\text{Fe}_{20}$, b) $\text{La}_{0.7}\text{Sr}_{0.3}\text{Mn}\text{O}_3$. The results of micromagnetic simulations are represented by dots and the theoretical calculation is displayed by a solid line. The color indicates the value of the bias magnetic field on which the calculation was performed.}
\end{figure}

\begin{figure}
\includegraphics[scale=0.6]{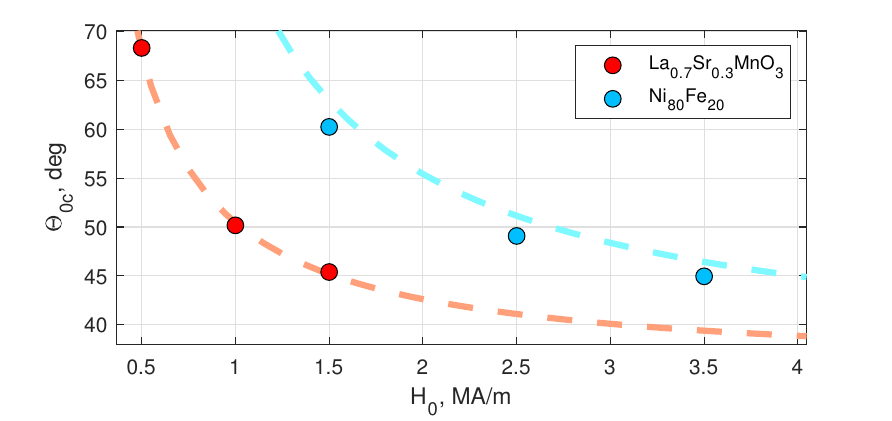}
\caption{\label{fig:FIG_3} The dependencies of the of the critical angle $\Theta_{0c}$ on the bias magnetic field $H_0$ for $\text{Ni}_{80}\text{Fe}_{20}$ and $\text{La}_{0.7}\text{Sr}_{0.3}\text{Mn}\text{O}_3$. The dots reflect the results of micromagnetic modeling and the dashed lines represent the numerical solution of the equation $N\left(\Theta_{0c}\right) = 0$.}
\end{figure}

The dependencies of the coefficient $N$ on the polar angle $\Theta_0$ of the bias magnetic field are shown in the Fig.~\ref{fig:FIG_2}. The theoretical dependence approximates well the simulation results for both materials. Moreover, it can be seen that the zero frequency shift coefficient is achievable at a certain angle $\Theta_{0}$. Let's call this angle critical $\Theta_{0c}$. From Fig.~\ref{fig:FIG_2} it can be seen that with an increase in the bias magnetic field $H_0$, critical angle moves towards smaller values. Using expression (\ref{eq:24}), we can only find the asymptotics of dependencies $\Theta_{0c}\left(H_0\right)$. To do this, it is enough to solve the equation $N\left(\Theta_{0c}\right) = 0$ assuming that the bias magnetic field $H_0$ is very large $H_0\gg M_s$. However, in the general case, there is no analytical solution for the equation $N\left(\Theta_{0c}\right) = 0$. Thus, it is impossible to get an analytical function $\Theta_{0c}\left(H_0\right)$. For an isotropic ferromagnet, we obtain an estimate of the limiting critical angle
\begin{equation}\label{eq:26}
   \Theta^i_{0c}=\frac{1}{2}\arccos\left(\frac{1}{3}\right)\approx35^0.
\end{equation}
For a ferromagnet in which there is only one easy axis, the formula for the smallest critical angle has the form
\begin{equation}\label{eq:27}
   \Theta^u_{0c}=\arccos\left(\frac{\sqrt{6}}{3}\sqrt{\frac{H_u+2Ms}{H_u\left(\cos(2\varphi_0+1)+2Ms\right)}}\right).
\end{equation}
If there are both an easy axis and a cubic crystallographic anisotropy in the material, then one can find an estimate for the critical angle $\Theta^c_{0c}$ by solving the biquadratic equation
\begin{equation}\label{eq:28}
   k_2q^4+k_1q^2+k_0=0,
\end{equation}
where
\begin{eqnarray}\label{eq:29}
q&=&\cos\left(\Theta^c_{0c}\right),\nonumber\\
k_0&=&-\frac{\gamma}{4}(18H_c-2H_u-4M_s),\nonumber\\
k_1&=&-3\gamma\left(\cos(\xi)^2-\frac{1}{2}\right)H_u\cos(\varphi_0)^2\nonumber\\
&-&\frac{3\gamma H_u}{4}\sin(2\xi)\sin(2\varphi_0)\nonumber\\
&+&\frac{3}{2}\gamma\left(H_u\cos(\xi)^2+15H_c-H_u-M_s\right),\nonumber\\
k_2&=&-\frac{45}{2}\gamma H_c(\cos(\varphi_0)^4-\cos(\varphi_0)^2+1).
\end{eqnarray}
In Fig.~\ref{fig:FIG_3} we show the dependencies of the critical angle on the bias magnetic field $H_0$ obtained by numerical solution of the equation $N\left(\Theta_{0c}\right) = 0$ and the results of micromagnetic simulations, which consists in the numerical solution of the Landau-Lifshitz equation (\ref{eq:01}). With the growth of the field $H_0$, the angle $\Theta_{0c}$ monotonously moves to a certain limit. For isotropic ferromagnetic thin films, this limit $\Theta^i_{0c}$ is determined by the expression (\ref{eq:26}) and does not depend on the saturation magnetization $M_s$. If the ferromagnet has an easy axis, then the angle $\Theta^u_{0c}$ differs from the limiting angle in the isotropic case $\Theta^i_{0c}$ and can be determined using the equation (\ref{eq:27}). The presence of cubic crystallographic anisotropy in the ferromagnetic film also leads to a difference between $\Theta^c_{0c}$ and $\Theta^i_{0c}$. The existence of such a limits can be explained as follows. From (\ref{eq:23}) it is clear that $N$ is a nonlinear coefficient in the energy expansion series near zero amplitude. This coefficient depends on the direction of magnetization corresponding to the minimum energy (\ref{eq:04}). With a sufficiently large bias magnetic field $H_0$, the direction of equilibrium magnetization $ \mathbf{e}_\zeta$ coincides with the direction of $\mathbf{H}_0$. Then, with an even greater increase in $H_0$, the direction of magnetization corresponding to the minimum energy no longer changes. This leads to the independence of the coefficient $N$ from the $H_0$ when $H_0\gg M_s$. That is why the $\Theta_{0c}\left(H_0\right)$ has a limit.

In conclusion, we theoretically demonstrated that achieving a zero nonlinear shift of a spin-transfer nanocillator frequency is possible using a bias magnetic field with a certain critical polar angle. We have shown that with an increase in the bias magnetic field, this angle moves towards smaller values. We have proved the existence of a minimum critical angle for both isotropic crystals and materials with uniaxial and cubic crystallographic anisotropies. The analysis has shown that the presence of uniaxial and cubic crystallographic anisotropies leads to a change in this minimum critical angle and the value of the critical angle at an arbitrary bias magnetic field compared with an isotropic material. The results of the investigation of the nonlinear frequency shift expand the prospects for the use of spin-transfer nanoscillators in spintronic devices.

The work was carried out within the framework of the state assignment of the Kotelnikov Institute of Radio Engineering and Electronics of the Russian Academy of Sciences.

 \section*{AUTHOR DECLARATIONS}
 \subsection*{Conflict of Interest}
 The authors have no conflicts to disclose.
 \subsection*{Author Contributions}
 {\bf Matveev Artem Andreevich:} Conceptualization (equal); Data curation (equal); Formal analysis (equal);
Investigation (equal); Methodology (equal); Software (equal); Validation (equal);
Visualization (equal); Writing – original draft (equal); Writing –
review \& editing (equal).
 {\bf Safin Ansar Rizaevich:} Conceptualization (equal); Data curation (equal); Formal analysis (equal); Funding acquisition (equal);
Investigation (equal); Methodology (equal); Supervision (equal); Validation (equal);
Visualization (equal); Writing –
review \& editing (equal).
 {\bf Nikitov Sergey Apollonovich:} Conceptualization (equal); Data curation (equal); Formal analysis (equal); Funding acquisition (equal);
Investigation (equal); Methodology (equal); Supervision (equal); Validation (equal);
Visualization (equal); Writing –
review \& editing (equal).
 
 \section*{Data Availability}
 The data that support the findings of this study are available
from the corresponding author upon reasonable request.
\nocite{*}
\bibliography{BIBLIO.bib}

\end{document}